\documentclass{aa}

\usepackage[varg]{txfonts}
\usepackage{graphicx}
\usepackage{multirow}
\usepackage{color}
\definecolor{blueAdobe}{RGB}{22,171,248}
\definecolor{yellowAdobe}{RGB}{255,161,0}
\definecolor{greenAdobe}{RGB}{27,167,9}
\definecolor{redAdobe}{RGB}{210,27,16}
\definecolor{violetPython}{RGB}{148, 0, 211}
\definecolor{brownPython}{RGB}{160, 82, 45}

\usepackage{natbib}
\bibpunct{(}{)}{;}{a}{}{,}

\begin{document}

\title{Comet formation in collapsing pebble clouds}
\subtitle{What cometary bulk density implies for the cloud mass and dust-to-ice ratio}

\author{
S.~Lorek\inst{\ref{inst1}}
\and B.~Gundlach\inst{\ref{inst2}}
\and P.~Lacerda\inst{\ref{inst1}}
\and J.~Blum\inst{\ref{inst2}}
}

\institute{
Max-Planck Institute for Solar System Research,
Justus-von-Liebig-Weg 3,
37077 G\"ottingen, Germany \\ \email{lorek@mps.mpg.de} \label{inst1}
\and Institut f\"ur Geophysik und extraterrestrische Physik, 
                Technische Universit\"at Braunschweig,
                Mendelssohnstr.~3,
                38106 Braunschweig, Germany \label{inst2}
}

\date{Received; Accepted}

\abstract
{Comets are remnants of the icy planetesimals that formed beyond the ice line in the Solar Nebula. Growing from $\mu\mathrm{m}$-sized dust and ice particles to $\mathrm{km}$-sized objects is, however, difficult because of growth barriers and time scale constraints. The gravitational collapse of pebble clouds that formed through the streaming instability may provide a suitable mechanism for comet formation.}
{
We study the collisional compression of silica, ice, and silica/ice-mixed pebbles during gravitational collapse of pebble clouds. Using the initial volume-filling factor and the dust-to-ice ratio of the pebbles as free parameters, we constrain the dust-to-ice mass ratio of the formed comet and the resulting volume-filling factor of the pebbles, depending on the cloud mass.
}
{
We use the representative particle approach, which is a Monte Carlo method, to follow cloud collapse and collisional evolution of an ensemble of ice, silica, and silica/ice-mixed pebbles. Therefore, we developed a collision model which takes the various collision properties of dust and ice into account. We study pebbles with a compact size of $1\,\mathrm{cm}$ and vary the initial volume-filling factors, $\phi_0$, ranging from $0.001$ to $0.4$. We consider mixed pebbles as having dust-to-ice ratios between $0.5$ and $10$. We investigate four typical cloud masses, $M$, between $2.6\times10^{14}$ (very low) and $2.6\times10^{23}\,\mathrm{g}$ (high).
}
{
Except for the very low-mass cloud ($M=2.6\times10^{14}\,\mathrm{g}$), silica pebbles are always compressed during the collapse and attain volume-filling factors in the range from $\langle\phi\rangle_V\approx0.22$ to $0.43$, regardless of $\phi_0$. Ice pebbles experience no significant compression in very low-mass clouds. They are compressed to values in the range $\langle\phi\rangle_V\approx0.11$ to $0.17$ in low- and intermediate-mass clouds ($M=2.6\times10^{17}-2.6\times10^{20}\,\mathrm{g}$); in high-mass clouds ($M=2.6\times10^{23}\,\mathrm{g}$), ice pebbles end up with $\langle\phi\rangle_V\approx0.23$. Mixed pebbles obtain filling factors in between the values for pure ice and pure silica. We find that the observed cometary density of $\sim0.5\,\mathrm{g}\,\mathrm{cm}^{-3}$ can only be explained by either intermediate- or high-mass clouds, regardless of $\phi_0$, and also by either very low- or low-mass clouds for initially compact pebbles. In any case, the dust-to-ice ratio must be in the range of between $3\lesssim\xi\lesssim9$ to match the observed bulk properties of comet nuclei.
}
{}


        \maketitle

        \section{Introduction}
        Comets are the remnants of icy planetesimals that formed beyond the snow line in the outer part of the Solar Nebula about $4.6\,\mathrm{Gyr}$ ago. They typically have  sizes from $1\,\mathrm{km}$ to $10\,\mathrm{km}$ \citep{Lamy2004} and consist of different ices (mainly water ice) and dust. The mean density of comet nuclei ranges from $0.4\pm0.2\,\mathrm{g}\,\mathrm{cm}^{-3}$ \citep{Ahearn2011} to $0.6\pm0.4\,\mathrm{g}\,\mathrm{cm}^{-3}$ \citep{Blum2006}. Recently, \citet{Sierks2015} determined the mean density of comet 67P/Churyumov-Gerasimenko at $0.470\pm0.045\,\mathrm{g}\,\mathrm{cm}^{-3}$. Comparing these estimates, $\rho_\mathrm{c}=0.5\pm0.3\,\mathrm{g}\,\mathrm{cm}^{-3}$ can be considered as a typical comet density. Therefore, to obtain such a low density, comets must be porous objects with a nucleus structure that can be thought of as a rubble-pile-like assemblage of dust and ice, held together by self-gravity rather than material strength \citep{Weissman2004}.

        Growth from $\mu\mathrm{m}$-sized dust and ice particles to $\mathrm{km}$-sized comets is difficult to explain because the process is affected by several growth barriers: the bouncing barrier for $\mathrm{mm}$- to $\mathrm{cm}$-sized particles \citep{Zsom2010}, the fragmentation barrier for $\mathrm{cm}$- to $\mathrm{m}$-sized bodies \citep{BlumWurm2008,Guettler2010}, and the radial drift barrier, which peaks for $\mathrm{m}$-sized objects at a heliocentric distance of $\sim1\,\mathrm{au}$ \citep{Weidenschilling1977}.

        One possible scenario for overcoming the growth barriers employs the bouncing barrier. A big particle sweeps up the small particles stuck at the bouncing barrier and thereby grows in mass-transfer collisions \citep{Windmark2012a,Drazkowska2013}. It was found that a $100\,\mathrm{m}$-sized body can form at a heliocentric distance of $3\,\mathrm{au}$ within $1\,\mathrm{Myr}$. However, the presence of big seed particles is necessary to trigger this mechanism. Among the main sources of collision velocities of small dust grains in the protoplanetary disk are Brownian motion and turbulence, which are stochastic processes. Thus, it is reasonable to assume that the collision velocities are distributed around a typical collision speed. This velocity distribution then helps to form those seeds,  because some particles fortuitously undergo only low-velocity sticking collisions and thus grow past the barriers \citep{Windmark2012b,Windmark2012c}. However, the dating of calcium-aluminium-rich inclusions (CAIs) and chondrules found in meteorites puts constraints on the timescale for the formation of planetesimals inside the snow line, which turns out to be a time span of only a few $\mathrm{Myr}$ after solar system formation \citep{Connelly2012}. For the cometesimals outside the snow line, the formation period is less constrained, but the lifetime of the gaseous disk of $\la10\,\mathrm{Myr}$ \citep{Fedele2010} is an upper limit. Therefore, the formation of $\mathrm{km}$-sized bodies via sweep-up growth is not fast enough, especially at large semi-major axes, where growth is generally slower than in the inner solar system.

        The streaming instability (SI) offers another mechanism for planetesimal formation. This type of two-fluid instability arises in the protoplanetary disk because dust and gas rotate at different velocities as a result of the sub-Keplerian speed of the gas and the non-perfect coupling of the pebbles to the gas in the disk. Using a linear perturbation analysis of the hydrodynamic equations, \citet{Youdin2005} showed that this situation is intrinsically unstable and tends to cluster dust particles, eventually leading to the formation of planetesimals. 

        The SI avoids the growth barriers by jumping directly from pebbles with Stokes number $\mathrm{St}\sim0.01-1$ (corresponding to a grain size of $\mathrm{mm}$ to $\mathrm{dm}$, depending on the radial location in the disk) to Ceres-sized planetesimals \citep{Johansen2007,BaiStone2010a,Drazkowska2014}. The SI is triggered once the local dust-to-gas ratio in the protoplanetary disk exceeds unity. In this case, the feedback of the dust on the gas motion via the drag force causes the local dust density enhancement to grow in a runaway process. This way, the local dust density grows up to $1000$ times the local gas density. If the density reaches Roche density, a gravitationally bound clump is formed. Subsequent gravitational collapse can finally form a planetesimal \citep{Johansen2007,Johansen2014,WahlbergJansson2014}.

        As argued by \citet{Skorov2012} and \citet{Blum2014}, the benefits of SI as a mechanism for forming comets is that it bypasses  growth barriers, while the collision speeds of pebbles in a collapsing cloud  also remain low, at most the escape velocity of the cloud. Hence, pebble fragmentation plays only a minor role, if at all, and the dominant collision-type is bouncing. The collapse of a cloud of bouncing pebbles then leads to a comet with a packing fraction of pebbles of $\phi_\mathrm{p}\approx0.6$. This packing fraction means that, assuming the comet is a sphere composed of spherical pebbles, only $60\,\%$ of the comet's volume is actually filled with pebbles; the remaining $40\,\%$ is void space. However, a packing fraction of $0.6$ assumes equal-sized pebbles. A size distribution changes the packing fraction, because the voids created by the big pebbles can be be filled with small pebbles, which results in a denser packing. The pebbles themselves are also not compact spheres, but possess a filling factor of $\phi\approx0.4$. As with the packing fraction, the volume-filling factor denotes that only $40\,\%$ of the pebble's volume is filled with solid material (ice or dust); again, the remaining $60\,\%$ is void space. This fluffiness is a result of the growth process in the protoplanetary disk. Hit-and-stick collisions of $\mu\mathrm{m}$-sized dust grains form fractal aggregates with low-filling factors. After reaching a certain size of typically a few $\mathrm{mm}$, the dust aggregates encounter the bouncing barrier and get compressed in bouncing collisions, which leads to a filling factor of $\sim0.4$ \citep{Zsom2010}. Experimental work by \citet{Weidling2009} and \citet{Guettler2009} also showed that the volume-filling factor of silica pebbles acquires a value of $\sim0.4$ after experiencing many bouncing collisions. The filling factor of the comet, $\phi_\mathrm{c}$, is then the product of both: $\phi_\mathrm{c}=\phi_\mathrm{p}\times\phi\approx0.24$ \citep{Skorov2012}. Comets are thus expected to be porous bodies, which agrees with recent density measurements and estimates \citep[e.g.][]{Blum2006,Ahearn2011,Sierks2015}.

        As previously mentioned, ice is a significant constituent of comets. Collision models based on laboratory experiments \citep{Guettler2010,Windmark2012a} so far only deal with silica aggregates, because experiments with ice particles are difficult to conduct. However, early results suggest that the threshold velocities for sticking, bouncing, and fragmentation of ice particles are by a factor of 10 higher than for silica \citep{Gundlach2011,Gundlach2015,Hill2015}. The compression curve of ice aggregates, on the other hand, cannot be scaled as easily as the threshold velocities (see Fig.~\ref{fig:compression_curves}).

        Here we describe numerical simulations of collapsing pebble clouds formed through the SI that are based on the energy formalism developed by \citet{WahlbergJansson2014}. We implement an improved collision model which allows us to use dust, ice, and dust/ice-mixed pebbles and to follow the filling factor of these pebbles. We conduct three sets of simulations, one with silica, one with ice, and one with mixed pebbles, to derive the compression behaviour of the pebbles during the collapse. Furthermore, we use the observed density of cometary nuclei and the estimated pebble packing to constrain the dust-to-ice ratio of the mixed pebbles, the cloud mass, and the pebble properties. 

        The paper is structured as follows: we briefly describe the laboratory experiments that lead to the compression curves of ice aggregates in Sect.~\ref{sec:experiments}. Then, we outline the numerical method in Sect.~\ref{sec:numericalmethod}. We present the simulation outcomes in Sect.~\ref{sec:results}. In Sect.~\ref{sec:discussion}, we discuss the results and the uncertainties. Finally, we conclude the paper by summarising our findings in Sect.~\ref{sec:conclusion}.

        \section{Laboratory experiments on dust and ice compression}
        \label{sec:experiments}

        Compression experiments on dust samples have been performed by \citet{Blum2004,Blum2006,Guettler2009} using silica particles as dust analogues. \citet{Guettler2009} approximate the compression curves by an analytic function of the form
\begin{align}
        \phi\left(p\right)=\phi_2-\frac{\phi_2-\phi_1}{\exp{\frac{\log_{10}{p}-\log_{10}{p_\mathrm{m}}}{\Delta}}+1}.
\label{eq:fermifunction}
\end{align}
Here, $p$ is the applied pressure, $p_\mathrm{m}$ is the pressure at which $50\,\%$ of the compression is reached, and $\Delta$ is the width of the transition regime. The factors,$\phi_1$ and $\phi_2$ are the minimum and the maximum volume-filling factors of the material, respectively. Depending on the nature of the compression experiments, i.e. unidirectional or omnidirectional, the maximum volume-filling factor varies between $\phi\approx0.3$ (unidirectional) and $\phi\approx0.6$ (omnidirectional). Table~\ref{tab:fitresults} summarises the results of their experiments on dust compression.

\begin{figure}
        \resizebox{\hsize}{!}{\includegraphics{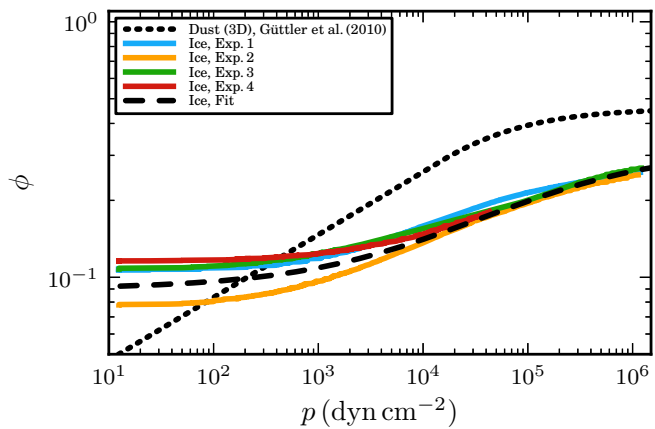}}
        \caption{Compression curves of the granular water ice samples consisting of micrometre-sized water ice particles (coloured curves). The initial volume-filling factor of the samples varies between $0.08$ and $0.12$. The dashed curve shows the best fit to the ice measurements. For comparison, the dotted line is the omnidirectional compression curve for silica particles used in the simulations \citep{Guettler2009,Guettler2010}.}
        \label{fig:compression_curves}
\end{figure}

        \citet{Guettler2009} give a value of $p_\mathrm{m}=1.3\times10^5\,\mathrm{dyn}\,\mathrm{cm}^{-2}$ for the static omnidirectional compression of the dust samples. To fit their dynamic compression experiments, they refined this value to $1.3\times10^4\,\mathrm{dyn}\,\mathrm{cm}^{-2}$ by comparing their measurements to the results of smoothed-particle hydrodynamics (SPH) simulations. \citet{Guettler2010} adopt the lower value in their model for the porosity evolution and furthermore extrapolate Eq.~\eqref{eq:fermifunction} for low pressures ($p<p_\mathrm{m}$) with a power-law function,
\begin{align}
        \phi\left(p\right)=\frac{\phi_1+\phi_2}{2}\cdot\left(\frac{p}{p_\mathrm{m}}\right)^{\frac{\phi_2-\phi_1}{\phi_2+\phi_1}\cdot\frac{1}{2\Delta\ln{10}}},
\label{eq:powerlaw}
\end{align}
to account for volume-filling factors lower than the ones used in their experiments.
Hence, we stick with the lower value of $p_\mathrm{m}$ and a compression curve for silica given by Eqs.~\eqref{eq:fermifunction} and \eqref{eq:powerlaw} (see dotted line in Fig.~\ref{fig:compression_curves}).

        In contrast to dust, the compression properties of granular water ice have, so far, not been measured. Thus, we performed laboratory experiments to investigate how the volume-filling factor of granular water ice changes when an external load is applied to the sample.

        We produce granular water ice samples by spraying liquid water into a cooled sedimentation chamber \citep[this method has been extensively discussed in previous works, see e.g.][]{Gundlach2011,Jost2013}. This method produces spherical micrometre-sized water ice particles with a mean radius of $1.45\,\mu\mathrm{m}$. After sedimentation of the micrometre-sized water ice particles in the sedimentation chamber, a porous water ice aggregate with an initial volume-filling factor of $\sim0.1$ is formed on a cold plate at the bottom of the chamber.

        These samples are then transported to a second experimental set-up that was constructed to measure the compressive strength of the water ice samples. Here, the samples are placed, together with the cold plate, inside a transport container filled with vaporised nitrogen to avoid sintering\footnote{Sintering normally starts to be significant (on the time scale of seconds to minutes) for temperatures $\ga180\,\mathrm{K}$. The experiments are carefully controlled to avoid temperatures above this critical value.} and condensation of frost on the sample's surface. The typical temperature of the cold plate during the transport process is $\sim100\,\mathrm{K}$ and the transfer time is less than $30\,\mathrm{s}$.

        To perform the compressive strength measurements, the transport container is placed on top of a scale and beneath a cooled piston (the temperature of the cold plate and the piston have never exceeded $125\,\mathrm{K}$ during the experiments). Then, the piston is slowly pressed into the granular sample by a stepper motor with a speed of $10\,\mu\mathrm{m}\,\mathrm{s}^{-1}$. The piston exerts a local compression on the sample, i.e. the material is able to flow into the non-compressed volumes of the sample.

        A total of four measurements are performed with this experimental set-up (solid curves in Fig.~\ref{fig:compression_curves}). The initial volume-filling factor varies between $0.08$ and $0.12$. Relatively low compressions ($< 10^3\,\mathrm{dyn}\,\mathrm{cm}^{-2}$) do not lead to an increase in the volume-filling factor of the water ice samples. However, higher compressions are sufficient to compress the granular water ice samples from the initial volume-filling factor to a volume-filling factor of $\sim 0.27$ (for a compression of $10^6\,\mathrm{dyn}\,\mathrm{cm}^{-2}$). The experiments stop at $10^6\,\mathrm{dyn}\,\mathrm{cm}^{-2}$, because the existing experimental set-up \citep[we adopted the experimental set-up that was used for the dust measurements by][]{Guettler2009} is not capable of detecting higher pressure values.

        For comparison, the compression curve of silica dust aggregates composed of micrometre-sized silica particles is shown by the dotted curve in Fig.~\ref{fig:compression_curves} \citep[see][for details]{Guettler2009,Guettler2010}. The comparison of the compression curves directly shows that the water ice aggregates require more load than the dust aggregates to reach the same volume-filling factor. This effect can be explained by the higher specific surface energy of water ice, compared to the silica dust particles \citep[see][]{Gundlach2011}. Compression of granular matter in this pressure regime is caused by the relocation of the particles in the sample by the rolling of the particles into the voids of the material, which then induces an increase of the mean coordination number per particle. Because of the higher specific surface energy of water ice, the rolling friction force of water ice is also enhanced in comparison to the silica dust particles, which implies that a higher pressure is required to rearrange the water ice particles in the samples. For a more detailed discussion of the compression physics of granular material, see \citet{Schraepler2015}.

        Since the granular ice aggregates are harder to compress, the experimental set-up is not capable of resolving the entire pressure curve. Thus, the pressure curves of the water ice samples do not saturate, as it is the case for comparable dust aggregates. Filling factors higher than $0.27$ might thus be possible for pressures exceeding $10^6\,\mathrm{dyn}\,\mathrm{cm}^{-2}$, but more experiments with a modified set-up would be required to test this and, thus, outside the objectives of this paper.

        To analyse the data in more detail, we apply Eq.~\eqref{eq:fermifunction} based on the Fermi distribution with logarithmic pressure to fit each individual data set. The values of the fit parameters are summarised in Table~\ref{tab:fitresults}. The application of the Fermi distribution is motivated by the fact that \citet{Guettler2009} use a similar function to fit the compression curves of the silica dust aggregates. To provide a general compression curve for granular water, we calculated the geometric mean of the individual fit parameters (see Table~\ref{tab:fitresults}).

        A comparison of the fit parameters of the water ice samples with the silica dust aggregates (unidirectional and omnidirectional, Table~\ref{tab:fitresults}) indicates that the water ice samples behave similarly to the unidirectional compression of the dust aggregates \citep[see][for details]{Guettler2009}. This means that the material is able to flow into the non-compressed areas of the samples. Only the parameter $\Delta$ is too high for fitting into this picture. This indicates that the compressional behaviour of water ice cannot only be explained by  shifting the compression curves of the dust aggregates to higher compression pressures.

\begin{table*}
\caption{Fit parameters for the silica and the granular water ice samples.}
\label{tab:fitresults}
\centering
    \begin{tabular}{llllll}
        \hline
        \hline
                Material & Compression & $\phi_1$ & $\phi_2$ & $p_\mathrm{m}$ & $\Delta$ \\
                & & & & $(\mathrm{dyn}\,\mathrm{cm}^{-2})$ & $(\mathrm{dex})$ \\
        \hline
                Silica & 1D & $0.15$ & $0.33$ & $5.6\times10^{4}$ & $0.33$ \\
                Silica & 3D & $0.12$ & $0.58$ & $1.3\times10^{4}$ & $0.58$ \\
                Ice, Exp.~1 & quasi-3D & $0.10\pm0.01$ & $0.27\pm0.01$ & $\left(3.2\pm0.1\right)\times10^{4}$ & $0.73\pm0.01$ \\
                Ice, Exp.~2 & quasi-3D & $0.07\pm0.01$ & $0.28\pm0.01$ & $\left(5.0\pm0.1\right)\times10^{4}$ & $0.83\pm0.01$ \\
                Ice, Exp.~3 & quasi-3D & $0.09\pm0.01$ & $0.45\pm0.01$ & $\left(109.7\pm0.5\right)\times10^{4}$ & $1.32\pm0.01$ \\
                Ice, Exp.~4 & quasi-3D &  $0.11\pm0.01$ & $0.31\pm0.01$ & $\left(14.0\pm0.3\right)\times10^{4}$ & $0.71\pm0.01$ \\
                \textbf{Ice, Geometric Mean} & quasi-3D & $0.09\pm0.01$ & $0.32\pm0.01$ & $\left(12.5\pm0.2\right)\times10^{4}$ & $0.87\pm0.01$ \\        \hline
        \end{tabular}
        \tablefoot{We adopted the values for silica from Table 1 in \citet[][]{Guettler2009}, with the refined value for $p_\mathrm{m}$ for the 3D compression. For ice, we give the best-fit parameters for each of the experiments shown in Fig.~\ref{fig:compression_curves}. The geometric mean yields the parameters for the general compression curve of ice. Unidirectional compression is 1D, omnidirectional compression is 3D. The fit parameters of ice aggregates are comparable to 1D ($\phi_1$, $\phi_2$, and $p_\mathrm{m}$) and 3D ($\Delta$) compression of silica; hence we label it quasi-3D (see Sect.~\ref{sec:experiments} for details).}
\end{table*}

        \section{Numerical method}
        \label{sec:numericalmethod}

        A comet of radius $5\,\mathrm{km}$ forming from ice pebbles with $1\,\mathrm{cm}$ radius, for example, would require $\sim10^{17}$ pebbles -- a number that is impossible to simulate directly. Therefore, we employ the representative particle approach, developed by \citet{Zsom2008}, for dust coagulation in the protoplanetary disk to model the collapse of the pebble cloud. The Monte Carlo approach is suitable because it allows modelling a large number, $N$, of physical particles by using only a small random sample of those particles. This means, $n\ll N$ particles out of the $N$ particles are randomly chosen, which are the representative particles. Thus, the fundamental assumption is that the distribution function of the representative particles can be regarded as the distribution function of the physical particles. The time evolution of the whole system is then given by following the change of the properties of the $n$ representative particles over time. The random sample can be small, which has the advantage of being computationally feasible, e.g. $100$ representative particles in contrast to the $10^{17}$ physical particles.

        \subsection{Cloud collapse model}
        \label{sec:model}
        The details of the collapse model are outlined in \citet{WahlbergJansson2014}. The model employs energy considerations for a pebble cloud that starts in virial equilibrium. Without any means of energy dissipation, the cloud remains in equilibrium because the pebbles move on orbits oscillating between apo- and pericentre. Elastic collisions only redistribute the orbital planes, which leads to random motion (pressure) counteracting gravity. The collapse sets in, because energy is dissipated in inelastic pebble collisions. The cloud contracts and releases gravitational energy to compensate for the dissipated energy. The collapse ends, once the cloud reaches the desired density at which all pebbles clump together and a solid body is formed. In our simulations, the collapse stops when the cloud reaches a density of $0.5\,\mathrm{g}\,\mathrm{cm}^{-3}$, which we consider here as the typical density of a comet.

        \subsection{Collision model for dust and ice pebbles}
        \label{sec:collisionmodel}
        To simulate collisions between pebbles, we use the collision model by \citet{Windmark2012a}. Based on the extensive work by \citet{Guettler2010}, \citet{Windmark2012a} identify four collision types as the most important for dust growth in the protoplanetary disk: sticking, bouncing, fragmentation, and partial fragmentation with mass transfer. This collision model is supported by laboratory experiments of colliding dust aggregates and provides mass and velocity dependent thresholds on the four collision regimes. Hence, it can be applied  not only to modelling the dust growth in the protoplanetary disk, but also to model the pebble collisions in the collapsing cloud.

        The threshold velocities for sticking, bouncing, and fragmentation of dust aggregates are given by \citet{Windmark2012a}:
\begin{align}
\Delta v_\mathrm{stick}^\mathrm{d}&=\left(\frac{m_\mathrm{p}}{m_\mathrm{s}}\right)^{-5/18}\,\mathrm{cm}\,\mathrm{s}^{-1}, \label{eq:thresholdsticking} \\
\Delta v_\mathrm{bounce}^\mathrm{d}&=\left(\frac{m_\mathrm{p}}{m_\mathrm{b}}\right)^{-5/18}\,\mathrm{cm}\,\mathrm{s}^{-1}, \label{eq:thresholdbouncing} \\
\Delta v_\mathrm{frag}^\mathrm{d}&=\left(\frac{m_\mathrm{p,t}}{m_{1.0}}\right)^{-0.16}\,\mathrm{cm}\,\mathrm{s}^{-1}. \label{eq:thresholdfragmentation}
\end{align}
where $m_\mathrm{p,t}$ is the mass of the projectile, or target, respectively. The normalising constants $m_\mathrm{s}=3.0\times10^{-12}\,\mathrm{g}$ and $m_\mathrm{b}=3.3\times10^{-3}\,\mathrm{g}$ are determined using laboratory data. The normalisation mass $m_{1.0}=3.67\times10^{7}\,\mathrm{g}$ describes the onset of fragmentation. By replacing $m_{1.0}$ with $m_{0.5}=9.49\times10^{11}\,\mathrm{g}$, one obtains the velocity threshold for which the largest remnant is half of the mass of the projectile or target. The fragmentation threshold velocity is calculated in the centre of mass reference frame. Thus, $\Delta v_\mathrm{frag}$ is compared to the centre of mass velocities of target and projectile, respectively:
\begin{align}
v_\mathrm{cms}^\mathrm{p}&=\frac{\Delta v}{1+m_\mathrm{p}/m_\mathrm{t}}, \label{eq:centreofmassvelocityprojectile} \\
v_\mathrm{cms}^\mathrm{t}&=\frac{\Delta v}{1+m_\mathrm{t}/m_\mathrm{p}}. \label{eq:centreofmassvelocitytarget}
\end{align}
For equal-mass pebbles, the centre of mass velocity is half of the relative velocity.

        Comets are very porous objects. Hence, we add the porosity evolution recipe for sticking and bouncing described in \citet{Guettler2010} to the collision model. Sticking collisions form fractal aggregates and the volume-filling factor decreases. \citet{Ormel2007} describe how to calculate the new filling factor of the resulting aggregate. In contrast, bouncing collisions compress the pebbles. The maximum compression per collision depends on collision velocity and pebble density, and thus on the pressure exerted on target and projectile. The compression curve gives the maximum filling factor that can be obtained for a certain pressure (see Fig.~\ref{fig:compression_curves}). The details are outlined in \citet{Guettler2009,Guettler2010} and \citet{Weidling2009}. However, for fragmentation and mass transfer, we simplistically assume no porosity change. 

        To include water ice pebbles into the simulations, we scale the threshold velocities Eqs.~\eqref{eq:thresholdsticking}-\eqref{eq:thresholdfragmentation} by a factor of ten to higher values, i.e. $\Delta v_\mathrm{stick}^\mathrm{i}=10\times\Delta v_\mathrm{stick}^\mathrm{d}$, and similarly for bouncing and fragmentation. This factor of ten is motivated by the fact that the specific surface energy and the sticking threshold velocity of water ice is also ten times higher \citep[][]{Gundlach2011,Gundlach2015,Hill2015}. The measured compression curve of ice shown in Fig.~\ref{fig:compression_curves} does not allow for an easy scaling. Thus, we use Eq.~\eqref{eq:fermifunction} with the mean values of the fit parameters given in Table~\ref{tab:fitresults}.

        \subsection{Collision model for dust/ice-mixed pebbles}
        \label{sec:collisionmodelmixed}
        Using pure ice and pure silica pebbles is a major simplification. There is evidence for $\mu\mathrm{m}$-sized pure water ice grains without refractory inclusions in the interior of comets \citep{Sunshine2007}. However, comet nuclei are also known to contain crystalline silicates, which cannot have formed in the cold outer part of the solar nebula \citep{McKeegan2006}. Thus, radial mixing must have taken place in the solar nebula \citep{BockeleeMorvan2002,Hanner2004,Cuzzi2006}. The coagulation timescale to grow from $\mu\mathrm{m}$-sized grains to $\mathrm{cm}$-sized pebbles is expected to be short \citep[$\la10^3\,\mathrm{yr}$,][]{Zsom2010} compared to the radial mixing timescale \citep[$\ga10^4\,\mathrm{yr}$,][]{BockeleeMorvan2002}. 

        Therefore, we assume that small silica and ice grains (monomers) originate inside and outside the snow line, respectively, and are mixed together in the comet-forming region. Thereby, the monomers coagulate to form $\mathrm{cm}$-sized dust/ice-mixed pebbles. Unfortunately, we do not know how mixed pebbles behave in collisions. The outcome will likely be in between the pure silica and the pure ice case, depending on the dust/ice content. We propose a simple interpolation scheme between both cases for mixed pebbles, which is based on the fractional abundances of dust and ice monomers (see Sect.~\ref{sec:interpolationscheme}).

        For our interpolation scheme to work, we assume a homogeneous mixture of silica and ice monomers within the mixed pebble. A pebble with a refractory core and an icy mantle is expected to behave like an icy pebble, but with the mass of a silica pebble, because the density ratio of silica to ice is $\sim3$. However, we do not consider core-mantle pebbles in our study.

        \subsubsection{Density of a mixed pebble}
        \label{sec:densitymixedpebble}
        A single pebble of mass $m$ is composed of $N_\mathrm{d}$ dust monomers of mass $m_{0,\mathrm{d}}$ and $N_\mathrm{i}$ ice monomers of mass $m_{0,\mathrm{i}}$. The total mass of dust and ice within the pebble is then $m_\mathrm{d}=N_\mathrm{d}\,m_{0,\mathrm{d}}$ and $m_\mathrm{i}=N_\mathrm{i}\,m_{0,\mathrm{i}}$, respectively, and the pebble composition is expressed in terms of the dust-to-ice ratio, $\xi\equiv m_\mathrm{d}/m_\mathrm{i}$. The dust-to-ice ratio is the mass ratio of dust to ice and not the volume ratio. Furthermore, we only consider water ice and neglect other ice species, such as carbon monoxide, carbon dioxide, methane, or ammonia.

        We first calculate the density of a mixed pebble with fixed dust-to-ice ratio $\xi$. The pebble mass is the sum of the masses of all dust and ice monomers:
\begin{align}
m&=m_\mathrm{d}+m_\mathrm{i} = \rho_\mathrm{d}V_\mathrm{d}+\rho_\mathrm{i}V_\mathrm{i},
\label{eq:pebblemass1}
\end{align}
where $V_\mathrm{d,i}$ are the compact volumes occupied by dust and ice monomers, respectively, and $\rho_\mathrm{d,i}$ are the corresponding monomer densities. On the other hand, the pebble mass is given by
\begin{align}
m=\frac{\rho_\bullet(V_\mathrm{d}+V_\mathrm{i})}{\phi},
\label{eq:pebblemass2}
\end{align}
where $\phi$ is the pebble-filling factor and $\rho_\bullet$ is the pebble density. Using $V=V_\mathrm{d}+V_\mathrm{i}$ and $\nu_\mathrm{d,i}=V_\mathrm{d,i}/V$, it follows that $\nu_\mathrm{d}+\nu_\mathrm{i}=1$. Together with the definition of $\xi$, Eqs. \eqref{eq:pebblemass1} and \eqref{eq:pebblemass2} can be solved for $\rho_\bullet$:
\begin{align}
\rho_\bullet=\phi\,\frac{\rho_\mathrm{d}\,\rho_\mathrm{i}(1+\xi)}{\xi\rho_\mathrm{i}+\rho_\mathrm{d}}=\phi\rho_\bullet^*, \label{eq:mixedpebbledensity}
\end{align}
where $\rho_\bullet^*$ is the compact density without any void space between the monomers. Thus, the density of the pebble depends on the dust-to-ice ratio. For $\xi=0$ (no dust), $\rho_\bullet^*=\rho_\mathrm{i}$ and for $\xi=\infty$ (no ice) we get $\rho_\bullet^*=\rho_\mathrm{d}$, as expected.

        \subsubsection{Interpolation scheme}
        \label{sec:interpolationscheme}
        To interpolate the collision behaviour between the two cases of pure dust and pure ice, we use the fractional abundance of dust monomers, $x=N_\mathrm{d}/(N_\mathrm{d}+N_\mathrm{i})$, because the thresholds for sticking, bouncing, and fragmentation depend on the ability of the collision to rearrange or break bonds between the monomers \citep{Dominik1997,Blum2000}. If there are more dust monomers than ice monomers, the pebble behaves more like a dust pebble. In the opposite case, when there are more ice monomers than dust monomers, the pebble behaves more like ice.

        Thus, we formulate the sticking threshold velocity for a mixed pebble as
\begin{align}
\Delta v_\mathrm{stick}^\mathrm{mixed} &= x\,\Delta v_\mathrm{stick}^\mathrm{d} + (1-x)\,\Delta v_\mathrm{stick}^\mathrm{i} \nonumber \\
&=(10-9x)\,\Delta v_\mathrm{stick}^\mathrm{d}.
\end{align}
Here, we use the fact that the threshold velocity for ice is ten times the value for dust (see Sect.~\ref{sec:collisionmodel}). Similar expressions can be derived for the bouncing and fragmentation threshold velocity.

        However, the remaining recipe of \citet{Windmark2012a} for determining masses of the remnant and fragments does not change in this case. However, there is one important assumption in the case of fragmentation: the composition of the pebble does not change. If,  prior to the collision, the pebble has a dust-to-ice ratio of $\xi$, the remnant and the fragments will have the same value for the dust-to-ice ratio after the collision. This should not cause any problems in fragmentation/coagulation cycles, unless pebbles are ground down to monomers. In this case, the composition can change as a result of the accretion of single monomers. The collision velocities in the collapsing pebble cloud, however, are too low for this effect to play an important role.

        Pebbles are compressed in bouncing collisions. The maximum compression is given by the compression curve $\phi(p)$ (see Eqs.~\eqref{eq:fermifunction} and \eqref{eq:powerlaw}), which provides the maximum volume-filling factor, $\phi$, that a pebble obtains for a given pressure $p$. As with the threshold velocities for sticking, bouncing, and fragmentation, the compression of a pebble is determined by rearranging bonds between monomers. Thus, we use the same linear interpolation scheme based on the fractional abundance of dust monomers, $x$, to get
\begin{align}
\phi^\mathrm{mixed}(p)=x\,\phi^\mathrm{d}(p) + (1-x)\,\phi^\mathrm{i}(p),
\end{align} 
where $\phi^\mathrm{d,i}(p)$ are the compression curves of pure silica and pure ice aggregates, respectively, which cannot be simply scaled as is the case for  threshold velocities (see Fig.~\ref{fig:compression_curves} and Table~\ref{tab:fitresults}).

        The change of the volume-filling factor in sticking collisions \citep[see][for details]{Ormel2007} is valid also for mixed pebbles. To compensate the suppression of fractal growth in collisions between aggregates and single particles, \citet{Ormel2007} introduce an additional term, $\Psi_\mathrm{add}\propto\exp(-\mu/m_\mathrm{F})$ ($\mu$ is the reduced mass of target and projectile), to their Eq.~(13). The mass scale $m_\mathrm{F}=10\,m_0$, which ensures the correction term vanishes for collisions between two massive aggregates, depends on the monomer mass $m_0$. For mixed pebbles, however, silica and ice monomers are present and thus we replace $m_0$ by the reduced monomer mass $m_0^{-1}\mapsto m_{0,\mathrm{d}}^{-1}+m_{0,\mathrm{i}}^{-1}$ in our implementation of the \citet{Ormel2007} porosity model.

        \subsection{Initial conditions}
        \label{sec:initialconditions}
        With this set-up (collapse and collision model), we conduct three sets of simulations. The initial compact pebble size, without void space, is $s^*=1\,\mathrm{cm}$. We vary the initial filling factor, $\phi_0$, of the pebbles and the pebble cloud mass, $M$. In the first set, we use water ice pebbles with a compact density of $\rho^*_\bullet=\rho_\mathrm{i}=1\,\mathrm{g}\,\mathrm{cm}^{-3}$. In the second set, we have silica pebbles with a compact density of $\rho^*_\bullet=\rho_\mathrm{d}=3\,\mathrm{g}\,\mathrm{cm}^{-3}$. In the third set, we simulate mixed pebbles with varying dust-to-ice ratios, $\xi$, and densities according to Eq.~\eqref{eq:mixedpebbledensity}.

        The cloud mass equals the mass of the final cometesimal, because it is assumed that the cloud collapses into one single object \citep{WahlbergJansson2014}. We use four different cloud masses equivalent to the mass of an object with radius $0.5\,\mathrm{km}$, $5\,\mathrm{km}$, $50\,\mathrm{km}$, and $500\,\mathrm{km}$, respectively, with a bulk density of $\rho_\mathrm{c}=0.5\,\mathrm{g}\,\mathrm{cm}^{-3}$, which is formed by the collapse of the cloud. We refer to these four masses as very low-, low-, intermediate-, and high-mass cloud or case, respectively. The exact masses of the clouds are $2.6\times10^{14}\,\mathrm{g}$ (very-low mass), $2.6\times10^{17}\,\mathrm{g}$ (low mass), $2.6\times10^{20}\,\mathrm{g}$ (intermediate mass), and $2.6\times10^{23}\,\mathrm{g}$ (high mass). The smallest size represents the size of a comet, whereas the largest one corresponds to the size of Ceres.

        We use initial filling factors in the range between $0.001-0.4$ to span the gap from very fluffy aggregates to compact pebbles that form when reaching the bouncing barrier. We do not expect filling factors much higher than $0.4$ because this is the limit given by the compression properties of the aggregates at the bouncing barrier \citep{Weidling2009,Zsom2010}. Very fluffy aggregates can form in hit-and-stick collisions of icy particles outside the snow line \citep[e.g.][]{Okuzumi2012,Kataoka2013}. Filling factors lower than $0.001$ are, in principle, possible in hit-and-stick collisions of ice \citep[][including collisional compression of the combined aggregate]{Okuzumi2012} but, due to the predominantly bouncing, we can always expect a certain amount of compression, which increases the filling factor rapidly.
Furthermore, in the simulations with mixed pebbles, we use four different initial filling factors ranging from $0.1$ to $0.4$ and six different dust-to-ice ratios between $0.5$ (ice-dominated) and $10$ (dust-dominated).
        We use a constant coefficient of restitution of $\approx0.7$ \citep{Blum1993, Weidling2012,Weidling2015}. However, changing the value of the coefficient of restitution slightly affects the final filling factor of the pebbles. We address this issue in the caveats.

        \subsection{Resolution study}
        \label{sec:resolutionstudy}

\begin{figure}
        \resizebox{\hsize}{!}{\includegraphics{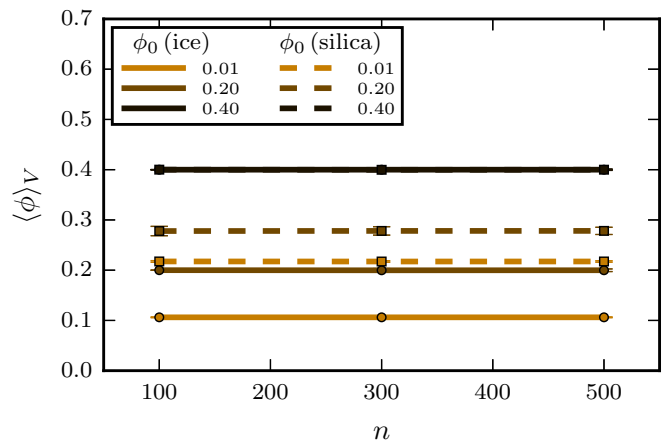}}
        \caption{Resolution study. Volume-weighted mean filling factor of pebbles as a function of the number of representative particles for different initial filling factors ($\phi_0=0.01$, $0.20$, and $0.40$) for the low-mass cloud. Increasing the number of particles does not change the final filling factor.}
        \label{fig:resolution}
\end{figure}
        We tested different numbers of representative particles and found no significant resolution dependency for the final volume-filling factor of the pebbles. Figure~\ref{fig:resolution} shows the volume-weighted mean filling factor of ice and silica pebbles for an increasing number of representative particles in the low-mass case. Increasing the number of particles does not change the final filling factor because each representative particle experiences, on average, $\ga 2000$ collisions. This is in accordance with \citet{Weidling2009} who reported saturation of the volume-filling factor after $\sim2000$ collisions. The mean number of collisions per representative particle also exceeds $2000$  in the intermediate- and high-mass cloud for $n=100$, which leads, together with the higher collision velocities in these cases, to a volume-filling factor approaching the maximum value \citep[$0.46$ for silica and $0.27$ for ice  according to][and Fig.~\ref{fig:vvf}]{Guettler2010}. Thus, we do not expect any changes when increasing the number of particles. However, there is some scatter in the final collapse time, which can be reduced by using more particles but, on average, we successfully reproduce the results found by \citet{WahlbergJansson2014}. Based on these tests, we use $n=100$ representative particles for the runs presented here.

        \section{Results}
        \label{sec:results}
        In this section, we present the results of our two sets of simulation runs. In Table~\ref{tab:simulationresultsfull}, we provide the values for the final filling factor of the pebbles obtained in our simulations as an overview.

        \subsection{Collision regimes}
        \label{sec:collisionregimes}
        We find that bouncing is the most frequent collision type in all four cloud masses, accounting for almost $100\,\%$ of pebble collisions. Some sticking occurs for ice, dust, and mixed pebbles alike in the very low-mass cloud and for small fragments in higher mass clouds.  Fragmentation and mass transfer contribute to the collision outcomes for higher cloud masses, which leads to a shift in the size distribution towards smaller pebbles in these clouds. However, the contribution by sticking, fragmentation, or mass transfer is insignificant compared to the very large number of bouncing collisions. Thus, for studying the porosity evolution of the pebbles, the a priori assumption of neglecting the porosity changes in fragmenting collisions in our collision model is justified.

        \subsection{Volume-filling factor}
        \label{sec:volumefillingfactor}

\begin{figure*}
        \resizebox{\hsize}{!}{\includegraphics{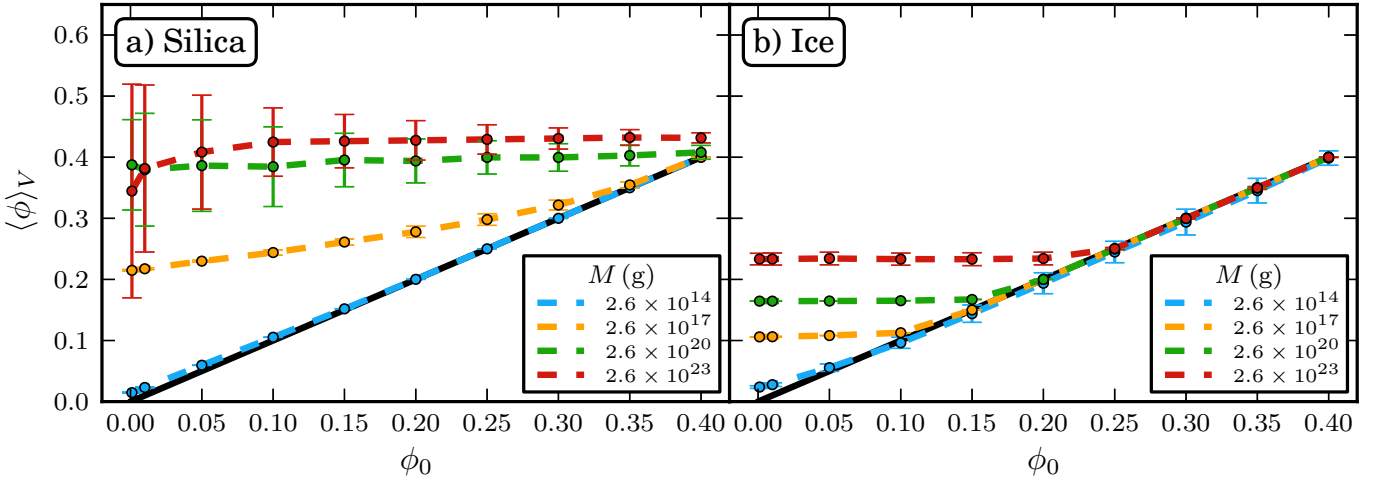}}
        \caption{Pebble compression for pure ice and silica pebbles. Volume-weighted mean filling factor at the end of collapse as a function of initial filling factor: (a) silica pebbles; (b) ice pebbles. The four cases of a very low-mass (\textcolor{blueAdobe}{blue}), a low-mass (\textcolor{yellowAdobe}{yellow}), an intermediate-mass (\textcolor{greenAdobe}{green}), and high-mass (\textcolor{redAdobe}{red}) cometesimal are shown. Along the black dotted line, the final filling factor equals the initial filling factor. The error bar indicate the standard deviation of the pebble ensemble. Silica pebbles are significantly compressed for $M>2.6\times10^{20}\,\mathrm{g}$. The maximum filling factor ice pebbles can obtain by compression is considerably lower than for silica, $\phi\sim0.23$ compared to $\phi\sim0.43$.}
        \label{fig:vvf}
\end{figure*}

\begin{figure*}
        \resizebox{\hsize}{!}{\includegraphics{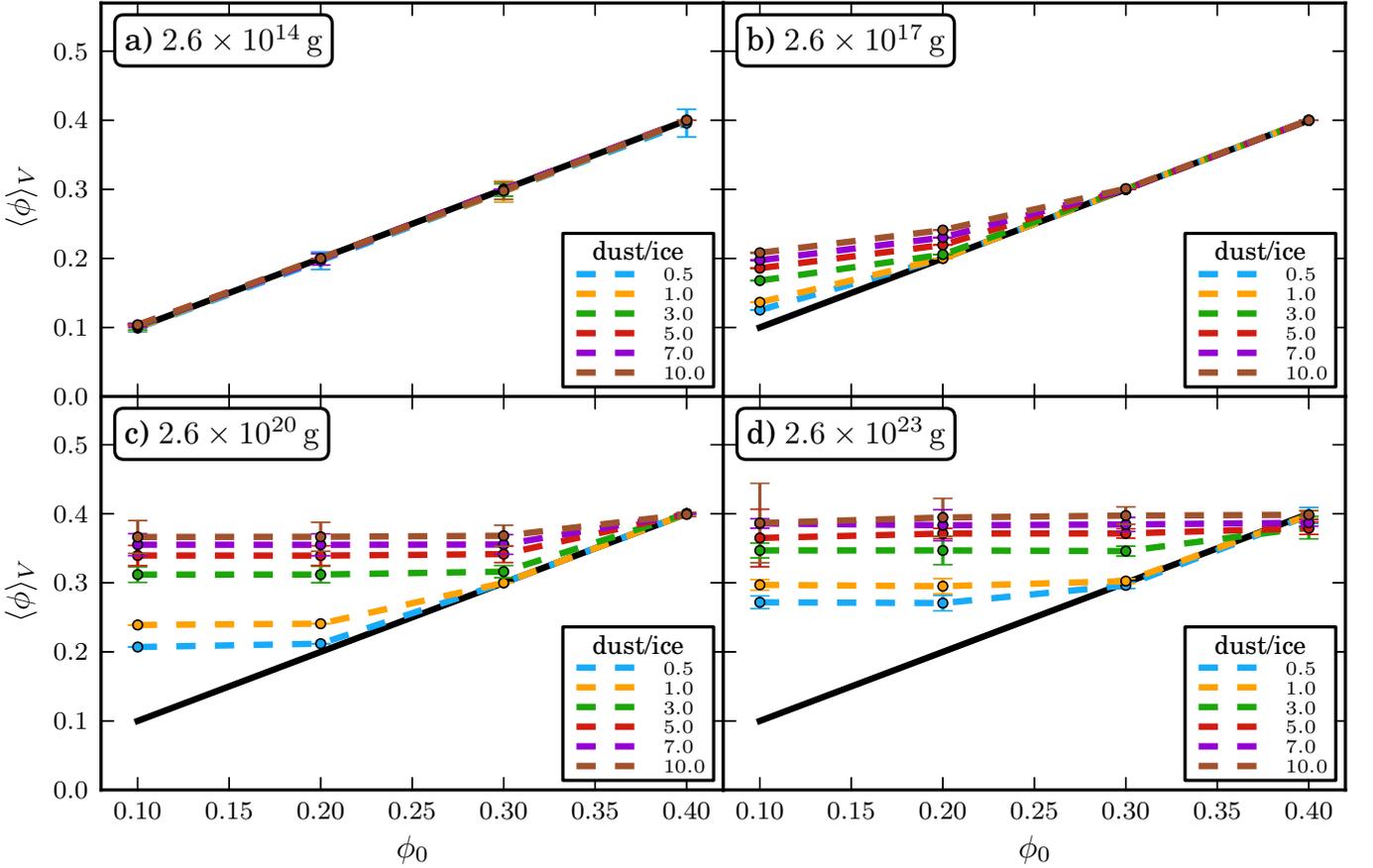}}
        \caption{Pebble compression for mixed pebbles. Volume-weighted mean filling factor at the end of collapse as a function of initial filling factor: (a) very low-mass cloud; (b) low-mass cloud; (c) intermediate-mass cloud; (d) high-mass cloud. Mixed pebbles with different dust-to-ice ratios are shown: $0.5$ (\textcolor{blueAdobe}{blue}, ice-dominated), $1.0$ (\textcolor{yellowAdobe}{yellow}), $3.0$ (\textcolor{greenAdobe}{green}, equal amount of dust and ice), $5.0$ (\textcolor{redAdobe}{red}), $7.0$ (\textcolor{violetPython}{violet}), and $10$ (\textcolor{brownPython}{brown}, dust-dominated). Along the black solid line, the final filling factor equals the initial filling factor. The error bar indicate the standard deviation of the pebble ensemble. As for pure silica or ice pebbles, compression increases with increasing cloud mass. Furthermore, compression increases with increasing dust-to-ice ratio. Pebbles with low dust-to-ice ratios behave like ice pebbles, while pebbles with high dust-to-ice ratios behave more like silica pebbles.}
        \label{fig:vvfmixed}
\end{figure*}

        We investigate the final volume-weighted mean filling factor of pebbles: 
\begin{align}
        \langle\phi\rangle_V=\frac{\sum_{i=1}^nN_iV_i\phi_i}{\sum_{i=1}^nN_iV_i}, \label{eq:volumeweightedmeanfillingfactor}
\end{align}
where $N_i=M/(n\,m_i)$ is the number of particles a representative particle represents, $V_i=V_i^*/\phi_i$ is the volume of a pebble ($V_i^*$ is the compact volume, i.e. the volume without void space), and $\phi_i$ is the volume-filling factor of a pebble. We weigh the filling factor with the volume, because we expect large and fluffy pebbles to dominate the  porosity of the comet.
        The mean value in Eq.~\eqref{eq:volumeweightedmeanfillingfactor} is a weighted sample mean of the pebble ensemble. To quantify the uncertainty of the volume-filling factor, we use the weighted sample standard deviation of the pebbles.

        \subsubsection{Pure dust and ice pebbles}
        \label{sec:volumefillingfactorpure}
        We find a clear difference in $\langle\phi\rangle_V$ for ice and silica and for different cloud masses. As can be seen in Fig.~\ref{fig:vvf}a for the very low-mass cloud, compression of silica pebbles is negligible and the initial filling factor is retained, unless rare sticking collisions decrease the filling factor by a small amount. Thus, the blue curve closely follows  the black line. In the low-, intermediate-, and high-mass clouds, silica pebbles are significantly compressed to aggregates with volume-filling factors in the range of $\langle\phi\rangle_V\approx0.22-0.43$, regardless of their initial filling factor. Compression increases with increasing cloud mass, but the pebbles approach the maximum compression of $\phi_\mathrm{max}\approx0.46$ \citep[][]{Weidling2009,Guettler2010} only in the intermediate- and high-mass clouds. The maximum value is reduced by a factor $0.79$ compared to $\phi_2$ in Table~\ref{tab:fitresults} because only the outer rim of the aggregate is compressed.

        Ice pebbles, in contrast, show a different behaviour (see Fig.~\ref{fig:vvf}b). As in the case of silica, compression of ice pebbles in the very-low mass cloud is negligible. However, even here a few sticking collisions reduce the filling factor slightly. In the low-mass cloud, pebbles are compressed to $\langle\phi\rangle_V\approx0.11$. In simulation runs starting with $\phi_0\ga0.11$, the ice pebbles retain their initial filling factor and the yellow line in Fig.~\ref{fig:vvf}b follows the black line. The same holds true for the intermediate- and high-mass clouds, but here the pebbles have at least $\langle\phi\rangle_V\approx0.16$, or approach their maximum filling factor of $\langle\phi\rangle_V\approx0.23$, according to the compression curve (see Fig.~\ref{fig:compression_curves}). Thus, $\phi=0.11$, $\phi=0.16$, and $\phi=0.23$ are lower boundaries for the compression of the ice pebbles. 

        \subsubsection{Mixed pebbles}
        \label{sec:volumefillingfactormixed}
        Figure~\ref{fig:vvfmixed} shows the compression of mixed pebbles with different dust-to-ice ratios in the four clouds. In the very low-mass cloud, pebbles retain their initial filling factor regardless of dust-to-ice ratio. This is consistent with the findings for the pure ice and silica pebbles, which correspond to dust-to-ice ratios equal to $0$ and $\infty$, respectively, in the very low-mass cloud (see Fig.~\ref{fig:vvf}). Pebble compression increases as the cloud mass increases from low to high. Additionally, increasing the dust-to-ice ratio also leads to an increased compression, because a higher dust content makes the pebbles weaker. The final filling factors are generally higher than for pure ice, but lower than for pure dust. In the intermediate- and high-mass clouds, the final filling factors are in the range $\langle\phi\rangle_V\approx0.21-0.40$, but only pebbles with a dust-to-ice ratio of $\xi\ga3-5$ obtain a filling factor close to $\langle\phi\rangle_V\approx0.40$.

        \subsection{Dust-to-ice ratio}
        \label{sec:dusttoiceratio}
        We relate the dust-to-ice ratio of the pebbles to the dust-to-ice ratio of the cometesimal and thereby constrain the initial mass of the pebble cloud in which the cometesimal has formed. This is possible because we assume that the cloud collapse results in a single object without pebbles being lost. Consequently, the dust-to-ice ratio of the final cometesimal is the same as the dust-to-ice ratio of a single pebble.

        The bulk density of the cometesimal is
\begin{align}
\rho_\mathrm{c}=\rho_\bullet^*\,\langle\phi\rangle_V\,\phi_\mathrm{P}, \label{eq:bulkdensitycometmixed}
\end{align}
where $\rho_\bullet^*$ is the compact density of the pebble, $\langle\phi\rangle_V$ is the final filling factor of the pebble, and $\phi_\mathrm{P}$ is the packing fraction of pebbles within in the cometesimal. According to \citet{Skorov2012}, the packing fraction is $\phi_\mathrm{P}\approx0.6$, which falls in the range between random loose packing (RLP) with $\phi_\mathrm{P}=0.55$ and random close packing (RCP) with $\phi_\mathrm{P}=0.64$.

        In our simulations, the bulk density of the cometesimal is a parameter and set to $\rho_\mathrm{c}=0.5\,\mathrm{g}\,\mathrm{cm}^{-3}$ \citep[e.g.][]{Sierks2015}. We use Eq.~\eqref{eq:bulkdensitycometmixed} to calculate $\phi_\mathrm{P}$ for the given compact density and final filling factor of the pebbles. The resulting value of $\phi_\mathrm{P}$ is then compared to the range given by RLP and RCP. This constrains the initial cloud mass because the final filling factor of the pebbles depends on the pebble cloud mass. Thus, initial conditions, i.e. $\xi$ and $M$, which are consistent with $\rho_\mathrm{c}$, are supposed to have $\phi_\mathrm{p}$ in the given range.

\begin{figure*}
        \centering
        \resizebox{\hsize}{!}{\includegraphics{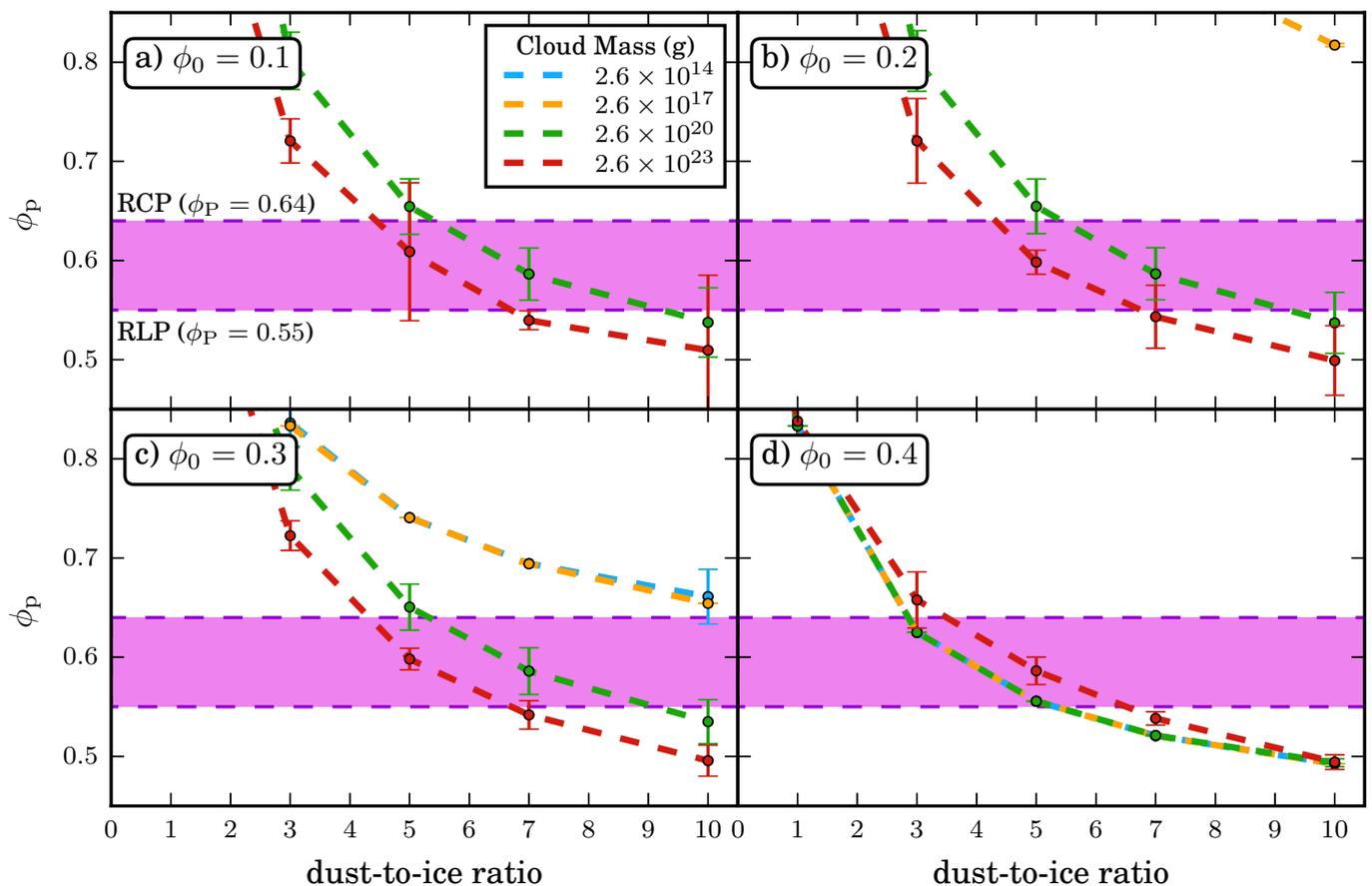}}
        \caption{Packing fraction of pebbles as a function of the dust-to-ice ratio. The four panels correspond to different initial filling factors for the pebbles: (a) of $\phi_0=0.1$; (b) $\phi_0=0.2$; (c) $\phi_0=0.3$; and (d) $\phi_0=0.4$. Each panel shows the results for the four different cloud masses as solid curves: very low-mass (\textcolor{blueAdobe}{blue}); low-mass (\textcolor{yellowAdobe}{yellow}); intermediate-mass (\textcolor{greenAdobe}{green}); and high-mass (\textcolor{redAdobe}{red}). In panels (a) and (b), the curves for the very low- and low-mass clouds are outside the plotting range, because the packing fraction becomes unphysically high ($\phi_\mathrm{P}\gtrsim1$). In panel (d), the green, blue, and yellow curves are overlapping because pebbles are barely compressed in these cases and thus have the same filling factor (see Fig.~\ref{fig:vvfmixed}). The shaded area (\textcolor{violetPython}{violet}) indicates the range of $\phi_\mathrm{P}$ given by the random loose packing (RLP, $\phi_\mathrm{P}=0.55$) and the random close packing (RCP, $\phi_\mathrm{P}=0.64$). The error bars show the uncertainty of $\phi_\mathrm{P}$, based on the uncertainty of the volume-filling factor of the pebbles. Initial conditions, i.e. dust-to-ice ratio and cloud mass, which are consistent with a cometary bulk density of $\rho_\mathrm{c}=0.5\,\mathrm{g}\,\mathrm{cm}^{-3}$, result in a value for $\phi_\mathrm{P}$ within the shaded area. Very low- and low mass clouds require initially compact pebbles ($\phi_0=0.4$) and dust-to-ice ratios $3\lesssim\xi\lesssim7$, whereas intermediate- and high-mass clouds need $3\lesssim\xi\lesssim9$, regardless of initial filling factor.}
        \label{fig:phip}
\end{figure*}

        In Fig.~\ref{fig:phip}, we plot the pebble packing fraction versus dust-to-ice ratio. In each panel, the solid lines connect the $\phi_\mathrm{P}$s calculated from the pebble-filling factors that were obtained in our simulations (see Table~\ref{tab:simulationresultsfull}). The error bars take the uncertainty of the volume-filling factor of the pebbles into account. Different line colours are for the different cloud masses. The shaded area highlights the range of $\phi_\mathrm{P}$ given by the limits of RLP and RCP. From Figs.~\ref{fig:phip}a and b, it is evident that initially porous pebbles in a very low- or low-mass cloud are not sufficiently compressed to be consistent with a cometary bulk density of $\rho_\mathrm{c}=0.5\,\mathrm{g}\,\mathrm{cm}^{-3}$ since they would require an unphysically dense packing ($\ga1$). Thus, the lines are out of range and not shown. An initial filling factor of $0.3$, however, requires a dust-to-ice ratio that is significantly higher than $10$ and dense packing for the very low- and low-mass clouds (Fig.~\ref{fig:phip}c). If the pebbles are initially compact ($\phi_0\ga0.4$), then very low- and low-mass clouds are also capable of reproducing the expected $\phi_\mathrm{P}$ for $3\lesssim\xi\lesssim7$ (see Fig.~\ref{fig:phip}d). The intermediate- and high-mass clouds lead to sufficiently compressed pebbles, regardless of their initial filling factor (see Figs.~\ref{fig:vvf} and \ref{fig:vvfmixed} and Table~\ref{tab:simulationresultsfull}). However, only pebbles with a dust-to-ice ratio in the range $3\lesssim\xi\lesssim9$ fall into the shaded region, which is consistent with a cometary bulk density of $0.5\,\mathrm{g}\,\mathrm{cm}^{-3}$. Although it is possible to constrain the dust-to-ice ratio, the cloud mass is still ambiguous: low-mass clouds and initially compact pebbles work as well as high-mass clouds with any initial pebble porosity.

\begin{table*}
\caption{Volume-weighted mean filling factor of pebbles for all runs performed in this study.}
\label{tab:simulationresultsfull}
\centering
\begin{tabular}{lcccccccccccc}
        \hline
        \hline
        $M$ &$\xi$ & $\rho_\bullet^*$ & \multicolumn{9}{c}{$\phi_0$} \\
        \cline{4-13}
        $(\mathrm{g})$ & & $(\mathrm{g}\,\mathrm{cm}^{-3})$ & 0.001 & 0.01 & 0.05 & 0.10 & 0.15 & 0.20 & 0.25 & 0.30 & 0.35 & 0.40 \\
        \hline
        \multirow{8}{*}{$2.6\times10^{14}$}  & 0 & 1.00 & 0.024 & 0.028 & 0.056 & 0.096 & 0.144 & 0.194 & 0.245 & 0.294 & 0.345 & 0.399 \\
        & 0.5 & 1.29 & \dots & \dots & \dots & 0.099 & \dots & 0.197 & \dots & 0.297 & \dots & 0.396 \\
        & 1.0 & 1.50 & \dots & \dots & \dots & 0.100 &  \dots& 0.200 & \dots & 0.297 & \dots & 0.400 \\
        & 3.0 & 2.00 & \dots & \dots & \dots & 0.101 & \dots & 0.200 & \dots & 0.299 & \dots & 0.400 \\
        & 5.0 & 2.25 & \dots & \dots & \dots & 0.103 & \dots & 0.200 & \dots & 0.300 & \dots & 0.400 \\
        & 7.0 & 2.40 & \dots & \dots & \dots & 0.103 & \dots & 0.199 & \dots & 0.300 & \dots & 0.400 \\
        & 10.0 & 2.54 & \dots & \dots & \dots & 0.104 & \dots & 0.200 & \dots & 0.298 & \dots & 0.400 \\
        & $\infty$ & 3.00 & 0.015 & 0.023 & 0.060 & 0.105 & 0.152 & 0.200 & 0.250 & 0.300 & 0.350 & 0.400 \\
        \hline
        \multirow{8}{*}{$2.6\times10^{17}$}  & 0 & 1.00 & 0.106 & 0.106 & 0.108 & 0.113 & 0.150 & 0.200 & 0.250 & 0.300 & 0.350 & 0.400 \\
        & 0.5 & 1.29 & \dots & \dots & \dots & 0.125 & \dots & 0.200 & \dots & 0.300 & \dots & 0.400 \\
        & 1.0 & 1.50 & \dots & \dots & \dots & 0.136 & \dots & 0.200 & \dots & 0.300 & \dots & 0.400 \\
        & 3.0 & 2.00 & \dots & \dots & \dots & 0.168 & \dots & 0.206 & \dots & 0.300 & \dots & 0.400 \\
        & 5.0 & 2.25 & \dots & \dots & \dots & 0.186 & \dots & 0.219 & \dots & 0.300 & \dots & 0.400 \\
        & 7.0 & 2.40 & \dots & \dots & \dots & 0.197 & \dots & 0.230 & \dots & 0.300 & \dots & 0.400 \\
        & 10.0 & 2.54 & \dots & \dots & \dots & 0.208 & \dots & 0.241 & \dots & 0.301 & \dots & 0.400 \\
        & $\infty$ & 3.00 & 0.215 & 0.217 & 0.230 & 0.244 & 0.261 & 0.278 & 0.298 & 0.322 & 0.355 & 0.400 \\
        \hline
        \multirow{8}{*}{$2.6\times10^{20}$}  & 0 & 1.00 & 0.164 & 0.164 & 0.165 & 0.165 & 0.167 & 0.200 & 0.250 & 0.300 & 0.350 & 0.400 \\
        & 0.5 & 1.29 & \dots & \dots & \dots & 0.207 & \dots & 0.212 & \dots & 0.300 & \dots & 0.400 \\
        & 1.0 & 1.50 & \dots & \dots & \dots & 0.239 & \dots & 0.241 & \dots & 0.300 & \dots & 0.400 \\
        & 3.0 & 2.00 & \dots & \dots & \dots & 0.312 & \dots & 0.312 & \dots & 0.316 & \dots & 0.400 \\
        & 5.0 & 2.25 & \dots & \dots & \dots & 0.340 & \dots & 0.339 & \dots & 0.342 & \dots & 0.400 \\
        & 7.0 & 2.40 & \dots & \dots & \dots & 0.355 & \dots & 0.355 & \dots & 0.356 & \dots & 0.400 \\
        & 10.0 & 2.54 & \dots & \dots & \dots & 0.366 & \dots & 0.367 & \dots & 0.368 & \dots & 0.399 \\
        & $\infty$ & 3.00 & 0.388 & 0.380 & 0.386 & 0.385 & 0.395 & 0.394 & 0.400 & 0.400 & 0.403 & 0.408 \\
        \hline
        \multirow{8}{*}{$2.6\times10^{23}$}  & 0 & 1.00 & 0.233 & 0.233 & 0.234 & 0.233 & 0.233 & 0.234 & 0.250 & 0.300 & 0.350 & 0.400 \\
        & 0.5 & 1.29 & \dots & \dots & \dots & 0.272 & \dots & 0.271 & \dots & 0.296 & \dots & 0.398 \\
        & 1.0 & 1.50 & \dots & \dots & \dots & 0.297 & \dots & 0.295 & \dots & 0.303 & \dots & 0.398 \\
        & 3.0 & 2.00 & \dots & \dots & \dots & 0.347 & \dots & 0.347 & \dots & 0.346 & \dots & 0.380 \\
        & 5.0 & 2.25 & \dots & \dots & \dots & 0.365 & \dots & 0.371 & \dots & 0.372 & \dots & 0.379 \\
        & 7.0 & 2.40 & \dots & \dots & \dots & 0.386 & \dots & 0.383 & \dots & 0.385 & \dots & 0.387 \\
        & 10.0 & 2.54 & \dots & \dots & \dots & 0.387 & \dots & 0.395 & \dots & 0.397 & \dots & 0.399 \\
        & $\infty$ & 3.00 & 0.345 & 0.382 & 0.408 & 0.425 & 0.426 & 0.428 & 0.429 & 0.431 & 0.432 & 0.432 \\
        \hline
\end{tabular}
\tablefoot{Columns from left to right: mass of the cometesimals, mass dust-to-ice ratio ($0$ for pure dust and $\infty$ for pure ice), density of the pebble, and final filling factors at the end of collapse. The initial compact pebble size is $1\,\mathrm{cm}$ in each simulation. $100$ representative particles were used. The typical uncertainties of the final filling factors are $\la15\,\%$. The individual uncertainties can reach 50\,\% in the high-mass cloud for pure dust pebbles with $\phi_0=0.001$, because of the presence of barely compressed pebbles.}
\end{table*}

        \section{Discussion}
        \label{sec:discussion}
        In this section we discuss the results presented in the previous section and we outline the caveats of our model.

        \subsection{Simulation results}
        \label{sec:simulationresults}

        \subsubsection{Pebble compression}
        Bouncing collisions account for almost $100\,\%$ of the collisional outcomes, regardless of the initial conditions (cloud mass and pebble filling factor) and material (see Sect.~\ref{sec:collisionregimes}). Collisional compression, therefore, determines the final value of $\langle\phi\rangle_V$. This can be explained by considering the typical collision velocity of pebbles, $v$, during collapse, which determines the compression pressure $p\propto\rho_\bullet v^2$ \citep{Guettler2010}. The virial velocity, which is a proxy for the velocity dispersion of the pebbles -- and hence the typical collision speeds -- is, by definition, $\sigma\approx\sqrt{GM/R}$, where $R$ is the instantaneous cloud radius, and $M$ is the (constant) cloud mass. Initially, the cloud radius equals the Hill radius, $R_0=R_\mathrm{H}=a\left(M/3M_\sun\right)^{1/3}$. We consider the cometesimal to be formed in the Edgeworth-Kuiper belt at a distance of $a=40\,\mathrm{au}$. The virial velocity at this heliocentric distance is then given by
\begin{align}   \sigma=2.9\,\left(\frac{a}{40\,\mathrm{AU}}\right)^{-1/2}\left(\frac{\rho_\mathrm{c}}{0.5\,\mathrm{g}\,\mathrm{cm}^{-3}}\right)^{1/3}\left(\frac{r}{5\,\mathrm{km}}\right)\,\eta^{-1/2}\,\mathrm{cm}\,\mathrm{s}^{-1}, \label{eq:virialvelocity}
\end{align}
where $r$ is the radius of the cometesimal in $\mathrm{km}$, and $\eta=R/R_0$ is the normalised size of the cloud \citep[see also][]{WahlbergJansson2014}.

        From Eq.~\eqref{eq:virialvelocity}, we infer that the typical collision velocities are of the order of $\mathrm{cm}\,\mathrm{s}^{-1}$ because $v\sim\sqrt{2}\sigma$. Thus, we mostly expect  bouncing collisions to occur \citep[see Fig.~7 of][]{Windmark2012a}. If we also take the higher threshold velocities for sticking, bouncing, and fragmentation of ice \citep[a factor 10 compared to silica, e.g.][]{Gundlach2015} into account, bouncing is the only relevant collision type because pebbles collide neither fast enough for fragmentation, nor slow enough for sticking, which agrees with our findings in Sect.~\ref{sec:collisionregimes}.

        As a consequence of frequent bouncing collisions, silica pebbles are compressed during the collapse towards a filling factor of $\phi\approx0.40$ \citep{Weidling2009,Guettler2010}. Thus, the silica pebbles are compact at the end of the collapse, even if they are porous at the beginning, and the cometary material consists of silica pebbles with a mean filling factor of $\langle\phi\rangle_V\approx0.40$ (see Fig.~\ref{fig:vvf}), which is in agreement with \citet{Skorov2012,Blum2014}. Ice pebbles, however, are harder to compress than silica pebbles and their maximum filling factor is lower (see Fig.~\ref{fig:compression_curves}). Therefore, for the same cloud mass, ice pebbles retain a higher porosity than silica pebbles, in agreement with Fig.~\ref{fig:vvf}. Increasing the cloud mass enhances the number of compact ice pebbles, because the velocity dispersion of pebbles, and thus the compression pressure, increases, as can be seen from Eq.~\eqref{eq:virialvelocity}.

        Ice pebbles keep their low filling factor throughout the collapse, unless they had been compressed already before (in the very low-, low- and intermediate mass cloud cases), or the cloud mass, $M$, is sufficiently high (see Fig.~\ref{fig:vvf}b). However, $500\,\mathrm{km}$ is approximately the size of Ceres and too large for a comet, unless the cloud fragments during the collapse, which produces a size distribution of objects. Comets would then represent the low-mass tail of this distribution. We intend to address this scenario in future work.

        Although the ice pebbles remain more porous than the silica pebbles, they do not retain an arbitrarily low filling factor. The lower threshold of $\phi$ is $\langle\phi\rangle_V\approx0.11$ for the low-mass cloud and $\langle\phi\rangle_V\approx0.16$ for the intermediate-mass cloud. This is plausible because the pressure determines the compression in a bouncing collision. For fixed pressure, a pebble can only be compressed if its filling factor is below the compression curve (see Fig.~\ref{fig:compression_curves}). Otherwise, the pebble is too compact and further compression requires a higher pressure. As argued above, higher compression pressure requires a higher cloud mass. Thus, only pebbles with a filling factor less than a certain threshold can be compressed, which explains why the lower boundaries given above shift to higher values for increasing cloud mass, as seen in Fig.~\ref{fig:vvf}b.
Mixed pebbles fall in between the limiting cases of pure ice and pure silica (see Fig.~\ref{fig:vvfmixed}).

        \subsubsection{Dust-to-ice ratio and comet density}
        Based on the assumptions of mixed pebbles and a cloud collapse into a single cometesimal, a dust-to-ice ratio in the range of between $3\lesssim\xi\lesssim9$ is necessary to be consistent with a cometary bulk density of $\rho_\mathrm{c}=0.5\,\mathrm{g}\,\mathrm{cm}^{-3}$ and a pebble packing of $\phi_\mathrm{P}\approx0.6$.

        From observations of cometary dust trails by the Infrared Astronomical Satellite (IRAS), \citet{Sykes1992} inferred a dust-to-volatile ratio of $\sim3$. The comet 9P/Tempel 1 was visited by the NASA \emph{Deep Impact} spacecraft and an impactor, launched from the spacecraft, hit the surface on 4 July 2005. From the ejected material, consisting of dust and water ice, \citet{Kueppers2005} estimated a dust-to-ice ratio of $>1$ for the nucleus of 9P/Tempel 1. However, the ejecta originated from a near-surface and ice depleted layer (the result of outgassing most likely), which was not representative for the bulk material of the comet. Thus, the ice proportion within the nucleus might still be higher, lowering the total dust-to-ice ratio. Based on GIADA (Grain Impact Analyser and Dust Accumulator) measurements aboard the \emph{Rosetta} spacecraft, \citet{Rotundi2015} have recently estimated the dust-to-gas ratio of the comet 67P/Churyumov-Gerasimenko as being $6\pm2$ for water ice only, and $4\pm2$ when $\mathrm{CO}$ and $\mathrm{CO}_2$ are taken into account. Our simulations are comparable with those measurements if comets form in pebble clouds of intermediate- or high-mass or in very low- and low-mass clouds, but with initially compact pebbles. This is because, in those cases, we find a dust-to-ice ratio of $>1$, which falls into the range $1<\xi<10$ for a typical cometary bulk density (see Fig.~\ref{fig:phip}). However, we can exclude  combinations of very low- and low-mass clouds with fluffy pebbles as the initial conditions for cometesimal formation, because dust-to-ice ratios of $\gg10$ are not observed in comets. Bodies with such high dust-to-ice ratios would resemble rocks or maybe rocky asteroids with only a very little amount of ice.

        With our simulations of either only ice or only silica pebbles we also cover two extreme cases of an ice and a silica cometesimal, respectively. The cloud contracts until it reaches a density of $0.5\,\mathrm{g}\,\mathrm{cm}^{-3}$. Thus, we can then check whether the values obtained for the mean filling factor of the pebbles allow for a cometesimal entirely composed of only one of the two materials by calculating the packing fraction according to \eqref{eq:bulkdensitycometmixed} for $\xi=0$ and $\xi=\infty$, respectively. In our simulations, the first case of only ice always produces  $\phi_\mathrm{P}\ga1$. This excludes pure ice cometesimals because the typical comet density of $0.5\,\mathrm{g}\,\mathrm{cm}^{-3}$ cannot be achieved with a packing fraction of $\phi_\mathrm{p}\approx0.6$ for the pebbles inside the comet \citep{Skorov2012,Blum2014} (see Fig.~\ref{fig:vvf}a). The second case with only silica pebbles yields packing fractions that are either too high ($\ga0.7$), or too low ($\la0.5$). However, in the very low- and low-mass clouds, $\phi_\mathrm{P}\approx0.6$ is possible for $0.15\lesssim\phi_0\lesssim0.3$. From a physical perspective, a comet consisting of only dust is unlikely, because volatiles like water ice are essential, e.g. to explain activity \citep[see][]{Blum2014,Blum2015}. On the other hand, only ice is also unlikely, because observations clearly show a certain amount of dusty material present in comets, e.g. dust trails or grains in the coma \citep{Sykes1992,Rotundi2015}. Furthermore, it can be argued that a comet density that is different to our choice of $0.5\,\mathrm{g}\,\mathrm{cm}^{-3}$ is also covered by recent density estimates of comets \citep[e.g.][]{Blum2006,Ahearn2011,Sierks2015} because all methods rely on certain assumptions, e.g. density estimates based on orbital changes that are due to non-gravitational forces require a detailed model of outgassing of volatiles. However, the density measurement of $\rho_\mathrm{c}=0.470\pm0.045\,\mathrm{g}\,\mathrm{cm}^{-3}$ \citep{Sierks2015} for the Jupiter family comet 67P/Churyumov-Gerasimenko by the \emph{Rosetta} spacecraft points towards a density near the value that is assumed here. Thus, our results imply that a mixture of ice and dust is necessary to match the observations. 

        \subsection{Cloud mass, pebble filling factor, and comet size}
        \label{sec:cloudmassandcometsize}
        Although our model constrains the dust-to-ice ratio required to match observations to a range between $3$ and $9$, the initial cloud mass and the initial pebble filling factor are ambiguous. Intermediate- and high-mass clouds, regardless of the initial pebble filling factor work as well as very low- and low-mass clouds with initially compact pebbles (see Sect.~\ref{sec:dusttoiceratio}). Therefore, it is necessary to discuss this result with respect to observed comet sizes and pebble formation in more detail. We know from observations that cometary nuclei typically have sizes ranging from $1\,\mathrm{km}$ to $10\,\mathrm{km}$ \citep{Lamy2004,Ahearn2011}. This size range is consistent with our model of cometesimal formation in very low- and low-mass clouds, which correspond to objects of size $0.5\,\mathrm{km}$ and $5\,\mathrm{km}$, assuming a bulk density of $0.5\,\mathrm{g}\,\mathrm{cm}^{-3}$, respectively. Thus, the cloud collapse directly leads to a comet. However, this requires initially compact pebbles. Starting with $\mu\mathrm{m}$-sized silica monomers, the growing aggregates encounter a bouncing barrier, which leads to the required compression \citep{Zsom2010}. However, based on the fact that ice is stickier, harder to fragment, and harder to compress \citep{Gundlach2011,Gundlach2015,Hill2015}, \citet{Okuzumi2012} showed that $0.1\,\mu\mathrm{m}$-sized ice monomers grow to very porous aggregates, which do not experience significant collisional compression. Thus, we leave the issue of whether or not compact icy aggregates can form for future studies.

        Our simulations also show that intermediate- and high-mass clouds are capable of reproducing the observed properties of cometary nuclei. The corresponding sizes of the cometesimals are $50\,\mathrm{km}$ and $500\,\mathrm{km}$, respectively. In these cases, the pebbles are not required to be initially compact as they are compressed during the collapse. However, compared to the observed comet sizes of $1\,\mathrm{km}$ to $10\,\mathrm{km}$, these cometesimals are too big. Thus, the cometesimal needs to fragment into several nuclei. One possibility enabling this to happen is during the collapse of the cloud by fragmentation into sub-clumps. The smaller sub-clumps then collapse into comet nuclei. Since the resulting objects are bound to the same cloud, binary formation should be possible \citep[][]{Nesvorny2010}. An alternative possibility is the fragmentation of cometesimals in mutual collisions of big cometesimals after the collapse \citep[][]{Morbidelli2015}. However, cometesimal collisions which are energetic enough to disrupt the colliding bodies can also be expected to significantly change the material properties, such as the density. As a consequence, the small fragments would neither resemble comets, nor would they carry a lot of information about the formation process of the cometesimal.

        In summary, for this work on comet formation: in the framework of the streaming instability followed by the gravitational collapse of pebble clouds, a dust-to-ice ratio exceeding unity is required to match the typical cometary bulk density of $0.5\,\mathrm{g}\,\mathrm{cm}^{-3}$, owing to the compression behaviour of the ice/dust-mixed pebbles during the collapse. This indicates the formation of cometesimals in intermediate- to high-mass clouds. However, this implies that either the cloud or the cometesimal needs to fragment to form $\mathrm{km}$-sized comet nuclei. Lower mass clouds are possible, if the pebbles already start off compact. However, very low- and low-mass clouds in combination with fluffy pebbles can be excluded as initial conditions for cometesimal formation because, in these cases, the pebble packing becomes unphysically high ($\phi_\mathrm{P}\ga1$). Considering the conditions in the protoplantetary disk, which probably produces compact mixed pebbles with a filling factor in the range $\phi\sim0.4$ for pure silica and $\phi\sim0.27$ for pure ice (see Fig.~\ref{fig:compression_curves}), our model predicts a dust-to-ice ratio of $\xi\sim3-9$ for comets. However, this needs to be investigated in more detail in future studies because it is not clear whether or not ice aggregates encounter a bouncing barrier in the way that silica particles do, while growing in the protoplanetary disk.

        \subsection{Caveats}
        \label{sec:caveats}
        We are aware of a number of caveats in our approach and try to draw some attention to these issues.

        \subsubsection{Pebble velocity distribution}
        First of all, as is done in \citet{WahlbergJansson2014}, we assume that the pebble speeds follow a Maxwellian velocity distribution. This is only true if the collisions are frequent and fully elastic (ideal gas). However, the key ingredient of the collapse model is the inelasticity of pebble collisions. Sticking collisions are fully inelastic, whereas bouncing collisions can be either fully elastic or dissipate a certain fraction of the collision energy, which is characterised by the coefficient of restitution. Therefore, the velocity distribution function cannot be Maxwellian. The pebbles behave like a granular gas, which is an ensemble of macroscopic particles evolving under inelastic collisions. It can be shown analytically that, for a homogeneous granular gas without external forces, the velocity distribution function is close to Maxwellian \citep{VanNoije1998}. The peak shifts slightly to lower velocities and the high-velocity tail is overpopulated because inelastic collisions tend to cool the system \citep{VanNoije1998}. However, gravity is important in the collapsing cloud and the density distribution of pebbles is not necessarily homogeneous, e.g. as a result of  clustering \citep{Goldhirsch1993}. Bearing this in mind, more work on the velocity distribution of pebbles has to be carried out before the assumption of a Maxwellian velocity distribution is revisited.

        \subsubsection{Pebble size and triggering the streaming instability}
        Another uncertainty of the model is the pebble size we use. The streaming instability is sensitive to the local metallicity $Z$ (i.e. the ratio of the dust and the gas column densities) of the protoplanetary disk and to the Stokes number of the pebbles. In simulations that do not take stratification of the disk that is due to the vertical component of the Sun's gravity into account, particles with $\mathrm{St}\sim1$ are concentrated most effectively. On the other hand, in stratified simulations, pebbles with $\mathrm{St}\sim0.3$ show a strong clustering behaviour \citep{Johansen2007,Johansen2014}. However, \citet{BaiStone2010a} reported clustering of pebbles via the streaming instability for $\mathrm{St}\ga0.01$, if $Z$ was significantly higher than the nominal value of the minimum mass solar nebula, $Z_\mathrm{MMSN}=0.01$ \citep{Hayashi1981}.

        The Stokes number depends on the properties of the gas and the pebbles. A minimum mass solar nebula model applied to $40\,\mathrm{au}$ shows that pebbles with radius $s\la800\,\mathrm{m}$ reside in the Epstein drag regime. The Stokes number can thus be written as
\begin{align}
        \mathrm{St}=0.7\times\,\left(\frac{\rho_\bullet}{3\,\mathrm{g}\,\mathrm{cm}^{-3}}\right)\,\left(\frac{s}{1\,\mathrm{cm}}\right)\,\left(\frac{a}{40\,\mathrm{au}}\right)^{3/2},
\end{align}
where $\rho_\bullet$ is the material density of the pebble, $s$ is the size in $\mathrm{cm}$, and $a$ is the semi-major axis in $\mathrm{au}$. Using the volume-filling factor, $\phi$, and indicating compact values with a $^*$, we obtain\begin{align}
        \mathrm{St}=0.7\times\phi^{2/3}\,\left(\frac{\rho_\bullet^*}{3\,\mathrm{g}\,\mathrm{cm}^{-3}}\right)\,\left(\frac{s^*}{1\,\mathrm{cm}}\right)\,\left(\frac{a}{40\,\mathrm{au}}\right)^{3/2}
\end{align}
because $\rho_\bullet\propto\phi$, and $s\propto\phi^{-1/3}$ by definition of the filling factor. In our simulations, we keep $\rho_\bullet^*$, $s^*$, and $a$ fixed, but vary $\phi$, which means $\mathrm{St}\propto0.7\times\phi^{2/3}$. We therefore cover a range in Stokes numbers with our simulations and for $\phi\la0.001$, both ice and silica pebbles have $\mathrm{St}\la0.01$, which is lower than the critical value found in \citet{BaiStone2010a}. Those pebbles would thus  not trigger the streaming instability. A Stokes number of $0.3$ \citep{Johansen2007,Johansen2014}, on the other hand, corresponds to a filling factor of $\sim0.28$ for silica. Thus, $\mathrm{cm}$-sized silica pebbles can trigger the SI, if they are compact as predicted by \citet{Zsom2010}. To include the possibility of very fluffy ice aggregates, as mentioned in \citet{Okuzumi2012} and \citet{Kataoka2013}, we have to take into account that those aggregates have to be very large to obtain a suitable Stokes number for streaming instability, e.g. $\mathrm{St}=0.3$ and $\phi=0.001$ requires $\mathrm{m}$-sized ice pebbles. Compact ice pebbles with $\phi=0.4$ of $\sim\mathrm{cm}$ in size, on the other hand, can also trigger the streaming instability. The same issues also apply  for mixed pebbles, because their densities fall in the range between pure ice ($1\,\mathrm{g}\,\mathrm{cm}^{-3}$) and pure silica ($3\,\mathrm{g}\,\mathrm{cm}^{-3}$). This problem of initial conditions has to be managed more rigorously in future studies.

        \subsubsection{Self-gravity of the cometesimal}
        In our model, pebbles are collisionally compressed during the collapse. Once the comet is formed, compression by self-gravity becomes significant. Assuming a comet of constant density in hydrostatic equilibrium, the central pressure can be estimated as
\begin{align}
        P_\mathrm{c}=\frac{2}{3}\pi G\rho_\mathrm{c}^2r^2,
\end{align}
where $G$ is the gravitational constant, $\rho_\mathrm{c}$ is the comet density, and $r$ is the comet radius \citep{Blum2006}. For a typical comet density of $\rho_\mathrm{c}=0.5\,\mathrm{g}\,\mathrm{cm}^{-3}$, a $0.5\,\mathrm{km}$-sized object has a central pressure of $P_\mathrm{c}\approx9\times10^{1}\,\mathrm{dyn}\,\mathrm{cm}^{-2}$, a $5\,\mathrm{km}$-sized objects has $P_\mathrm{c}\approx9\times10^{3}\,\mathrm{dyn}\,\mathrm{cm}^{-2}$, a $50\,\mathrm{km}$-sized objects already has  $P_\mathrm{c}\approx9\times10^{5}\,\mathrm{dyn}\,\mathrm{cm}^{-2}$, and a $500\,\mathrm{km}$-sized body has an even higher central pressure of $9\times10^{7}\,\mathrm{dyn}\,\mathrm{cm}^{-2}$. Comparing these values to the compression curves for ice and silica pebbles shown in Fig.~\ref{fig:compression_curves}, we see that gravitational compression indeed plays a role for objects $r\ga5\,\mathrm{km}$. Consequently, the filling factors found for the pebbles must be regarded as lower limits. This also includes  the mixed pebbles, because the pure silica and ice cases are limiting cases for dust-to-ice ratios of $0$ and $\infty$, respectively.

        Furthermore, we implicitly assume that the cometesimals are primordial objects. Further evolution of these icy bodies, like collisions between cometesimals or mass loss owing to activity, is excluded in our analysis. Although, those processes would arguably only affect the upper layer of the body.

        \subsubsection{Collision model for mixed pebbles}
        We constructed a collision model for mixed pebbles by interpolating physical properties, such as sticking, fragmentation, or compression of the pebbles between the two cases of pure ice and pure silica. The interpolation scheme is based on the fractional abundance of silica monomers within the pebble, because the collision outcome and the compression depends on the ability of the collision to rearrange or break monomer bonds \citep[see Sect.~\ref{sec:interpolationscheme} and][]{Dominik1997,Blum2000}. Thus, the approach is suited for pebbles that consist of a homogeneous mixture of $\mu\mathrm{m}$-sized silica and ice monomers. This is a reasonable assumption, because of the stochastic coagulation process in the solar nebula. In contrast, pebbles with a refractory core and an icy mantle are assumed to behave like ice pebbles but with enhanced mass, which is mainly determined by the silica core. However, we do not consider core-mantle pebbles here.

        As described in Sect.~\ref{sec:interpolationscheme}, our approach does not take compositional changes of the pebble into account. This simplification breaks down if the collisions become energetic enough to grind down pebbles into monomers. The following accretion of single monomers, which are either silica or ice, leads to aggregates with a different dust-to-ice ratio to  the progenitor pebbles. However, the collision velocities in the collapsing pebble cloud are not high enough for this to be significant. In agreement with \citet{WahlbergJansson2014}, we observe  collision velocities, which are close to the virial velocity, in the less massive clouds. The velocities are of the order of a few $\mathrm{cm}\,\mathrm{s}^{-1}$ (see Eq.~\eqref{eq:virialvelocity}), which lead to bouncing collisions only. In more massive clouds, the velocities are initially higher due to the higher virial velocity, which are $1-2\,\mathrm{m}\,\mathrm{s}^{-1}$, see Eq.~\eqref{eq:virialvelocity}. However, the rapid energy dissipation, which is due to the onset of fragmentation in this velocity range (see Eqs.~\eqref{eq:thresholdfragmentation}, \eqref{eq:centreofmassvelocityprojectile}, and \eqref{eq:centreofmassvelocitytarget}), leads to velocities significantly lower than virial velocity again and also lower than the fragmentation threshold. Thus, pebbles and small fragments again undergo bouncing collisions. For decreasing dust-to-ice ratios, i.e. increasing ice content, the threshold velocities shift to even higher values and catastrophic fragmentation of pebbles is even less likely. However, for other applications, where velocities are higher and a significant amount of monomers is produced, a recipe for changing the composition of the aggregate might be necessary.

        Furthermore, the compression of the pebbles also has to be performed with care. We use the omnidirectional dynamical compression curve for silica pebbles given by \citet{Guettler2010} (see dotted line in Fig.~\ref{fig:compression_curves}). However, for the ice aggregates, it remains inconclusive whether the compression measured in the laboratory experiments is unidirectional or omnidirectional (see Sect.~\ref{sec:experiments}). Furthermore, the prescription for the porosity evolution given in \citet{Weidling2009} and \citet{Guettler2010} is tailored to fit silica particles. Adapting this model for ice by exchanging the compression curve only, is clearly not satisfactory. The same applies to mixed pebbles, where we combined the collision model of silica and ice. Therefore, we need more laboratory experiments for developing a model of ice compression in bouncing collisions, which is as detailed as the silica model.

        \subsubsection{Coefficient of restitution of pebbles}
        In our simulations, we chose a value of $\sim0.7$ for the coefficient of restitution, which is suitable for dust aggregates and grazing collisions \citep{Blum1993,Weidling2012}. \citet{Weidling2015} performed microgravity experiments with $\mathrm{mm}$-sized dust aggregates and found that a coefficient of restitution uniformly distributed in a range between $0.29$ and $0.81$ with a mean value of $0.55$ explains their experiments best. On the other hand, microgravity experiments with $\mathrm{mm}$- to $\mathrm{cm}$-sized ice pebbles have shown that the values for the coefficient of restitution are uniformly distributed between $0.06$ and $0.84$ with a mean value of $0.45$ \citep{Heiselmann2010} or between $0.08$ and $0.65$ with a mean value of $0.36$ \citep{Hill2015}. Thus, our choice of coefficient of restitution is in agreement with both types of materials. 

        However, changing the coefficient of restitution slightly affects the final filling factor of the pebbles. A lower value increases the amount of energy dissipated in a single collision. As a consequence, the cloud collapse proceeds faster, the collision velocities are lower, and also the total number of collisions is lower. Thus, the pebbles are compressed less. From Eq.~\eqref{eq:bulkdensitycometmixed} we see that a lower filling factor of the pebbles increases the necessary packing fraction of the pebbles in the comet to obtain a comet density of $0.5\,\mathrm{g}\,\mathrm{cm}^{-3}$. Hence, the curves in Fig.~\ref{fig:phip} shift slightly upwards and the range of dust-to-ice ratios shifts to higher values. However, this  effect has no significant influence on our results for the very low- and low-mass clouds, because the pebbles must be compact already at the beginning. In the intermediate- and high-mass clouds, the coefficient of restitution effect is partially compensated for by the higher collision velocities owing to the higher cloud mass, which leads to stronger compression. However, since pebble compression is determined by rearranging monomers, a lower coefficient of restitution should lead to a higher pebble compression. Our collision model does not capture this effect, because the pebble compression is determined independently from the coefficient of restitution, which only enters  the collision velocity.  Hence, we plan to investigate the influence of the coefficient of restitution in more detail in future work.

        \section{Conclusion}
        \label{sec:conclusion}
        We investigated the collapse of a pebble cloud based on the model developed by \citet{WahlbergJansson2014} and followed the porosity evolution of the pebbles by using the collision model of \citet{Windmark2012a} for dust, which we adapted for ice by scaling the threshold velocities up by a factor 10 \citep[see e.g.][]{Gundlach2015}. Furthermore, a recipe for the evolution of the filling factor of dust pebbles in sticking and bouncing collisions from \citet{Guettler2010} was used. In addition, we developed an interpolated collision model for mixed pebbles containing both ice and dust. We also present novel results from laboratory experiments measuring the compressive strength of aggregates composed of $\mu\mathrm{m}$-sized water ice particles. These measurements are used to derive the evolution of the filling factor of the silica, ice, and mixed pebbles during the gravitational collapse. 

        We tested four initial cloud masses ranging from very low ($2.6\times10^{14}\,\mathrm{g}$) to high ($2.6\times10^{23}\,\mathrm{g}$), corresponding to cometesimals with radii varying from $0.5\,\mathrm{km}$ to $500\,\mathrm{km}$ and a bulk density of $0.5\,\mathrm{g}\,\mathrm{cm}^{-3}$. We start with $1\,\mathrm{cm}$-sized pebbles that possess initial filling factors of between $\phi_0=0.001$ (very porous) and $\phi_0=0.4$ (compact), and investigated how the filling factor changes during the collapse, depending on the cloud mass. We conducted three sets of simulations, one with only silica pebbles, a second one with only ice pebbles, and a third one with mixed pebbles with varying dust-to-ice ratios. Additionally, we calculated the pebble packing required to produce the typical cometary bulk density of $0.5\,\mathrm{g}\,\mathrm{cm}^{-3}$ and investigate how this depends on the cloud mass, initial filling factor, and dust-to-ice ratio (see Fig.~\ref{fig:phip}). 

        Our main findings can be summarised as follows:
\begin{itemize}
        \item The laboratory measurements of the compressive strength of granular ice show that water ice aggregates are harder to compress than silica aggregates (see Fig.~\ref{fig:compression_curves}). This results in a lower volume-filling factor for the ice aggregates compared with silica aggregates, if compressed with the same pressure.
        \item Except for the very low-mass cloud, silica pebbles are always compressed and obtain volume-filling factors in the range of $\langle\phi\rangle_V\approx0.22-0.43$ at the end of the collapse, regardless of $\phi_0$ (see Fig.~\ref{fig:vvf}a).
        \item Ice pebbles experience no significant compression in the very low-mass cloud. They are compressed to $\langle\phi\rangle_V\approx0.11$ and $\langle\phi\rangle_V\approx0.17$ in low- and intermediate-mass clouds, respectively; in high-mass clouds, ice pebbles end up with $\langle\phi\rangle_V\approx0.23$ (see Fig.~\ref{fig:vvf}b). These values are lower boundaries for the pebble volume-filling factor. If the pebbles are more compact at the beginning, they retain their initial filling factor.
        \item Mixed pebbles obtain filling factors in between the values for pure ice and pure silica.
        \item Pebble compression, whereby the pebble packing is in agreement with the limits given by random loose packing ($\phi_\mathrm{P}=0.55$) and random close packing ($\phi_\mathrm{P}=0.64$) for a cometesimal with a bulk density of $\rho_\mathrm{c}=0.5\,\mathrm{g}\,\mathrm{cm}^{-3}$, can be achieved in two indistinguishable ways:
\begin{itemize}
\item in very low- and low-mass clouds for initially compact pebbles,
\item and in intermediate- and high-mass clouds, regardless of the initial filling factor of the pebbles.
\end{itemize}
The latter implies that one cometesimal forms more than one comet nucleus. In any case, a dust-to-ice ratio in the range $3\lesssim\xi\lesssim9$ is necessary to match the observed bulk properties of comet nuclei.
\end{itemize}

\begin{acknowledgements}
We thank Karl Wahlberg Jansson for fruitful discussions. The ice compression experiments were performed under DLR grant 50WM1236. We thank Tobias Eckhardt for his help in the laboratory. We also thank the editor, the anonymous referees, and the language editor for their comments, which helped to improve the quality of this manuscript. S.~Lorek is a member of the International Max-Planck Research School for Solar System Science at the Max-Planck Institute for Solar System Research, G\"ottingen.
\end{acknowledgements}

        \bibliographystyle{aa}
        \bibliography{references}

\end{document}